\documentclass{article}
\begin{document}
\title{The zero point field in low light level experiments.
\author{Jacques Moret-Bailly 
\footnote{Laboratoire de physique, Université de Bourgogne, BP 47870, F-21078 Dijon cedex, France. 
email : Jacques.Moret-Bailly@u-bourgogne.fr}}}
\maketitle

\begin{abstract}
The existence of the zero point electromagnetic field in the dark is a trivial property of classical 
electromagnetism. Splitting the total, genuine electromagnetic field into the sum of a conventional field and a 
zero point field is physically meaningless. If a receiver attenuates the genuine field down to a zero 
conventional field, it remains a zero point field having the coherence and the phase of the conventional field, 
and vice-versa for the amplification of the zero point field by a source.

Nonlinear optical effects must be written using the genuine field, so that at low light levels they become 
linear in relation to the conventional field. The result of the interpretation of all observations, even at low 
light levels, is exactly the same in quantum electrodynamics and in the semi-classical theory.

The zero point field is stochastic only far from the sources and the receivers; elsewhere, it is shaped by 
matter, it may be studied through fields visible before an absorption or after an amplification.

Two examples are given : the computation of fourth order interferences and the use of the impulsive 
stimulated Raman scattering with ordinary incoherent light in astrophysics.

pacs{42.25Bs, 42.50Gy}

\end{abstract}

\maketitle % %********************** Section 1 
%________________________________________________________________

\section{Introduction.}
The electromagnetic field radiated by an oscillating dipole, studied by Hertz in the nineteenth century, gives 
a particular solution of Maxwell's equations with a singular point at the origin. A linear development of this 
solution using solutions regular at the origin requires an infinite number of solutions. More physically, adding 
the solution for the dipole and another solution, the total energy may be higher, equal or lower than in the 
second solution, that is the dipole emits, refracts or absorbs energy; the last case corresponds to a partial 
cancellation of the exciting field by the field emitted by the dipole; this superposition of waves shows that 
while an oscillating dipole refracts or absorbs a part of the energy radiated by an other dipole, it diffracts a 
large fraction of this energy. Both demonstrations show that, in a semi-classical theory, the absorption by 
atoms of the light emitted by other atoms is slow, it remains a residual field named now the zero point field. 
While the existence of this field is trivial, the computation of the mean value of its intensity $h\nu/2$ per 
monochromatic mode, by Planck and Nernst \cite{Planck,Nernst}, was difficult.

The oscillating dipole is a particular system of moving charges; any moving electron radiates a field, but if it 
belongs to a stationary system, in the average, it does not radiate energy: Sommerfeld's electron does not 
fall on the proton, but the fluctuations of its perturbations by the zero point field produce the Lamb shift 
\cite{Power}.

The building of the zero point field shows that it is an ordinary field; thus it may be amplified by a source; 
the starts of the laser pulses show that a spontaneous emission is exactly the amplification of the zero point 
field.

In this paper we qualify the electromagnetic field as follows :

-''genuine'' for the total electromagnetic field;

''zero point" for the field which remains after the largest, coherent, physically possible absorption of the 
genuine field;

''conventional'' for the genuine minus the zero point;

''stochastic'' for a stochastic zero point field.

\medskip

For a long time, the sensitivity of the detectors of light was bad so that the zero point field is neglected in 
the conventional classical electrodynamics; in modern optics, the ''genuine field'' $\hat E$ is the sum of the 
''conventional field'' $E$ and the zero point field $E_0$; consequently the sensitivity of the photosensors is 
revisited  in section 2. A consequence  is the correct classical computation of fourth order interferences in 
section 3.

The semiclassical emission or absorption is modelled by an excitation of a mono- or polyatomic molecule 
by an electromagnetic field up to a barrier between the two involved relative minimums of potential $W_1$ 
and $W_2$; if the initial and final states are stationary $|W_2 - W_1| = h\nu$ \footnote{As a 
monochromatic function is physically meaningless, we should write $\int{\frac{e(\nu){\rm 
d}\nu}{\nu}}=h$, with $|W_2 - W_1|=\int{e(\nu){\rm d}\nu}$} . But our macroscopic experiments 
usually use nearly plane modes while the molecules emit light generally through dipoles or quadrupoles. 
Quantum mechanics transforms the geometry of the waves using the ''reduction of the wave packet''. 
Section 4 gives the classical reduction of the wave packet.

''Stochastic electrodynamics" is often used to avoid confusing the genuine with the conventional classical 
electrodynamics. Section 5 shows that this name is not very convenient because while the zero point field is 
stochastic far from the sources and the absorbers, this field is shaped by the sources and the absorbers.

The semi-classical theory is often preferred to quantum electrodynamics in nonlinear optics because it 
appears simpler and more concrete. The equivalence of the results provided by both theories, showed by 
Marshall \& Santos \cite{Marshall} disappears only if the zero point field is neglected. As an example, 
section 6 describes a light matter interaction whole ignorance by the astrophysicists seems a consequence 
of the difficult use of quantum electrodynamics, while the right classical electrodynamics leads easily to a 
result which could avoid to look for astonishing objects in the universe.
\section{Absorption and detection.} % Section 2
Usually, we write that the intensity absorbed or detected by a photocell is proportional to the square of the 
amplitude of the conventional electric field, this square being proportional to the conventional flux of 
electromagnetic energy. This supposes that there is no coherence between the conventional field and the 
zero point field, an assumption which is false because a spontaneous emission is an amplification of a mode 
of the zero point field (see section 4). The equations of absorption must fulfil two conditions: i) be written 
using only genuine fields (which may be zero point fields); ii) preserve the zero point field in the average. 
As a zero point field included in the genuine field is absorbed, an equivalent field must be reemitted (or the 
genuine field is attenuated down to the zero point field). Remark that in the dark, cold, good photocells 
generate a noise signal which seems produced by the fluctuations of the zero point field.

The net available energy on a receiver is proportional to 
\begin{equation} 
\hat E^2-E_0^2=(\beta E_0)^2-E_0^2=2(\beta- 1)E_0^2+((\beta-1)E_0)^2.
\end{equation} 
Usually $E_0^2$ is neglected, the usual rule is got; on the contrary, supposing that $\beta$ is nearly one, 
$((\beta-1)E_0)^2$ may be neglected; for a given optical configuration, the time- average of the zero point 
amplitude is constant, so that the detected signal is proportional to $(\beta-1)E_0$ that is to the amplitude 
of the conventional field, with $E_0$ as reference of phase. With an incoherent source, the phase factor 
fluctuates: without a sophisticated detection nothing appears.

\section {Fourth order interferences.}%Section 3

A sophisticated detection is performed in the fourth order interference experiments with photon counting: 
two elementary measurements are done while the phase is constant (see, for instance, 
\cite{Clauser,Gosh,Ou1,Ou2,Kiess}). The result of these experiments is easily got {\it qualitatively} using 
the classical rules \cite{M942}, but the contrast of the computed fringes is lower than shown by the 
experiments. In the simplest experiment \cite{Gosh} two small photoelectric cells are put in the 
interference fringes produced by two point sources; the interferences are not visible because they depend 
on the fast changing difference of phase $\phi$ of the modes of the zero point field amplified by the 
sources. The sources are weak; the signal is the correlation of the counts of the cells.

Distinguishing the photoelectric cells by an index j equal to 1 or 2, set $\delta_j$ the difference of paths for 
the light received by the cells. The amplitude of the conventional field received by a cell is proportional to 
$\cos (\pi\delta_j/\lambda+\phi/2)$, so that, assuming the linearity in the conventional field, the probability 
of a simultaneous detection is proportional to 
\begin{equation} 
\cos (\frac{\pi\delta_1}{\lambda}+\frac{\phi}{2})\cos (\frac{\pi\delta_2}{\lambda}+\frac{\phi}{2}).
\end{equation} 
The mean value of this probability got by an integration over $\phi$ is zero for $\delta_1-
\delta_2=\lambda/2$, so that the visibility has the right value 1. Assuming the usual response of the cells 
proportional to the square of the conventional field, the visibility would have the wrong value 1/2 
\cite{Mandel}.
\section{Classical reduction of the wave packet.}%Section 4

The reduction of the wave packet breaks the symmetry of the waves, transforming, in particular, a local 
wave into a beam, for instance a dipolar wave into a plane wave.

The polarisation of a transparent matter by a light beam may be observed by a variation of the energy 
levels, or detecting Kerr effect,...Thus the beginning of a pulse of light must transfer energy to matter, and 
this energy is recovered in its tail (except for a small incoherent Rayleigh scattering). Thus, in the tail, the 
field is amplified, although there is no transition, no inversion of population, the polarisation mixing only 
slightly the initial state of the molecules with other states. This power of amplification applies not only to the 
exciting mode, so that many modes, usually initially at the zero point, are amplified, later reabsorbed in the 
medium : there is an dynamical equilibrium between the exciting field, the other modes and the polarisation 
of the molecules. The modes which are excited are dipolar or quadrupolar, they radiate far only the small 
incoherent Rayleigh field : they may be qualified ''local''. On the contrary, the interactions of the light pulse 
are strong because they are coherent.

As the local modes are amplified, the strongest and longest fluctuations of their field may be able to excite 
molecules up to a barrier, such that an absorbing transition occurs; the mean energy of the local field, then 
of the molecules is decreased, the amplification tail of the pulse is decreased, the medium has absorbed the 
light. In a laser, a similar process explains the coherent amplification by incoherently pumped molecules.
\section{Is the zero point field stochastic ?}%Section 5

Whichever their far or close origin, the electromagnetic fields are mixed, so that splitting the 
electromagnetic field into the zero point field and the conventional field is physically meaningless; this 
splitting is an anticipation of the interaction of the field with a receiver able to extract from the mode all 
available energy. A pure, stochastic zero point field exists only far from sources, in the dark, coming from 
many far incoherent sources. On the contrary, the field emitted by a molecule which radiates or absorbs 
energy is, near the molecule, mostly the field of an oscillating dipole or quadrupole; even if an absorption 
re-establishes the mean value of the zero point field, the field is shaped by the molecule near it, not 
stochastic. Using the results of the previous section, this property extends to the modes of collimated 
beams; if a sheet of dark glass reduces the conventional intensity of a beam to zero, the phase of the beam 
remains written in the remaining zero point field.

A macroscopic consequence of the structuring of the zero point field by matter is observed in the Casimir 
effect \cite{Casimir}: long wavelengths are rejected from inside two parallel plates, so that a lower 
pressure of radiation attracts the plates.

The emission of a field during the absorption of a quantum of energy is not instantaneous; during this 
emission and a short time after it, the probability for an extraordinarily strong and long fluctuation of the 
field is lowered, so that a sub-poissonian photon statistic appears; neglecting the space-time structuring of 
the field leads to a poissonian statistic \cite{Short, Glauber}. 

Squeezing is also an elementary observation of a non-stochastic zero point field.

\section{Low level "Impulsive Stimulated Raman Scattering" (ISRS).}%section 6
Quantum electrodynamics leads to consider the photon not only as an amount of energy but as a particle 
too; although this last concept is sometimes considered as dangerous and rejected \cite{WLamb}, this 
concept led the astrophysicists to reject light-matter interactions as an alternative to the Doppler effect to 
explain observed redshifts of far objects. It is a rejection of coherent interactions.

ISRS, known since 1968 \cite{Giordmaine} is now commonly used \cite{Yan,Nelson}. It is not a simple 
Raman scattering, but a parametric effect, combination of two {\it space-coherent} Raman scattering, so 
that the state of the interacting molecules is not changed. The hot \footnote{The temperature of a spectral 
line is deduced from Planck's laws} exciting beam and its scattered beams interfere into a single frequency, 
redshifted, beam; the cold beam is blueshifted.

 ISRS is obtained using ultrashort light pulses, that is "pulses shorter than all relevant time constants" 
\cite{GLamb}, usually femtosecond laser pulses. As it has no intensity threshold it works with the light 
pulses which make the ordinary incoherent light. The relevant time constants in a gas may be adapted to 
incoherent light :

i) to avoid that the collisions destroy the coherence of the excitation of the molecules, the pressure must be 
very low;

ii) to obtain an interference of the scattered and incident lights into a single frequency light, the period which 
corresponds to the virtual Raman transition must be larger than the length of the impulsions. The molecules 
must have transitions in the radiofrequencies, generally hyperfine transitions.

Decreasing the intensity of the beams down to a value for which $\hat E$ is not much higher than $E_0$, 
the Raman scattered conventional amplitude $A$ proportional to the intensity writes

\begin{equation} 
A \propto\hat E^2-E_0^2=(\beta E_0)^2-E_0^2=2(\beta- 1)E_0^2+((\beta-1)E_0)^2.
\end{equation} 
The last term may be neglected, the first, proportional to the incident amplitude, represents the usual 
spontaneous coherent Raman amplitude; the interference reduces the incident and scattered wave vectors 
into a single wave vector, so that the scattered light is not on a cone as in usual laser coherent Raman 
experiments.

The Universe, provides good experimental conditions for a confusion of this interaction with a Doppler 
effect: the paths are long and the pressures often low, a lot of mono- or polyatomic molecules, perturbed 
by Zeeman effect which are observed have hyperfine structures. The absorption spectra of the molecules 
which are destroyed at their first collision, H$_2^+$ for instance cannot be seen because the redshift of 
their absorption spectra widens, thus weakens, their lines.

\medskip
Similarly, all optical effects become linear at low light levels.

\section{Conclusion.}
 The zero point field, often, improperly, qualified ''stochastic'' is a trivial component of the genuine classical 
electromagnetic field. An artificial subtraction of this field leads to the conventional classical 
electromagnetism, breaking the equivalence of the classical and quantum electrodynamics.

Many authors tried to demonstrate that the semi-classical theory is not correct at low light levels; their 
demonstrations use the conventional classical electrodynamics which is an approximation of the genuine 
classical electrodynamics; this approximation fails at low light levels.

Quantum and classical electrodynamics have their specific advantages : Quantum electrodynamics provides 
ready to use properties or postulates, but a common improper use of some of its concepts, the photon for 
instance, leads to wrong conclusions \cite{WLamb}; classical electrodynamics is more intuitive, but it 
requires often more complicated demonstrations. It is so common and useful to use both theories than 
some physicists think that the zero point field is introduced by quantum electrodynamics !

The teachers should replace unnecessarily approximate rules, for instance the first Planck's law, by the 
rigorous rules.


\begin{thebibliography}{25}
\bibitem{Planck} Planck, M., 1911, {\it Verh. Deutsch. Phys. Ges,} {\bf 13}, 138
\bibitem{Nernst} Nernst, W., 1916, {\it Verh. Deutsch. Phys. Ges,} {\bf 18}, 83
\bibitem{Power} Power, E. A., 1966, {\it Am. J. Phys.} {\bf 34} 516
\bibitem{Marshall} Marshall, T. W. \& E. Santos, 1988 {\it Found. Phys}, {\bf 18} 185; 1989, {\it Phys. 
Rev.} {\bf A 39}, 6271
\bibitem{Clauser} Clauser J. F. , Horne M. A. , Shimony A. \& Holt R. A. , 1969, {\it Phys. Rev. Lett. 
},{\bf 23}, 880
\bibitem{Gosh} Gosh R. \& Mandel L. , 1987, {\it Phys. Rev. Lett.} ,{\bf 59}, 1903 
\bibitem{Ou1} Ou Z. Y. \& Mandel L. , 1988, {\it Phys. Rev. Lett.} , {\bf 61}, 54
\bibitem{Ou2} Ou Z. Y. \& Mandel L. , 1990, {\it J. Opt. Soc. Am.} , {\bf 7}, 2127
\bibitem{Kiess} Kiess T. E. , Shih Y. H. , Sergienko A. V. \& Alley C. O. , 1993, {\it Phys. Rev. Lett.} , 
{\bf 71}, 3893
\bibitem{M942} Moret-Bailly J. , 1994, {\it J. Optics,}  {\bf 25}, 263
\bibitem{Mandel} Mandel L, 1983, {\it Phys. Rev.}  {\bf 28}, 929
\bibitem{Casimir} Casimir, H. B. G., 1948, {\it Proc. K. Ned. Akad. Wet.}, {\bf 51}, 793
\bibitem{Short} Short R. \& Mandel L. 1983, {\it Phys. Rev. Lett.} {\bf 51}, 384
\bibitem{Glauber} Glauber R. J. 1966. {in Physics of Quantum Electronics}, Kelley P. L. et al. ed. 
McGraw-Hill, New York, 788
\bibitem{WLamb} Lamb W. E. Jr., 1995, {\it Appl. Phys.}, {\bf B60}, 77
\bibitem{Giordmaine}Giordmaine, J. A. , M. A. Duguay \& J. W.Hansen, 1968, IEEE J. Quantum 
Electron., 4, 252
\bibitem{Yan} Yan Y.-X., Gamble E. B. Jr. \& Nelson K. A. , 1985, {\it J. Chem Phys. } {\bf 83}, 5391
\bibitem{Nelson} Nelson K. A. \& Fayer M. D., 1980, {\it J. Chem. Phys} {\bf 72}, 5202
\bibitem{GLamb} Lamb, G. L. Jr., 1971, {\it Rev. Mod.Phys} {\bf 43 }, 99
\bibitem{Moret} Moret-Bailly J. 1998, {\it Quantum and Semiclassical Optics} {\bf 10}, L35
\bibitem{Moret2} Moret-Bailly J. 2001, {\it J. Quantit. Spectr. \& Radiative Ttransfer} {\bf 68}, 575

\end{thebibliography}
\end{document}